\documentclass[twocolumn,showpacs,preprintnumbers,amsmath,amssymb]{revtex4}

\usepackage{graphicx}
\usepackage{dcolumn}
\usepackage{bm}

\begin{document}


\title{Frustration-induced valence bond crystal and its melting in Mo$_3$Sb$_7$}

\author{T. Koyama,$^{1}$ H. Yamashita,$^{1}$ Y. Takahashi,$^{1}$ T. Kohara,$^{1}$ I. Watanabe,$^{2}$ Y. Tabata,$^{3}$ and H. Nakamura$^{3}$}
\affiliation{$^{1}$Graduate School of Material Science, University of Hyogo, Kamigori, Ako-gun, Hyogo 678-1297, Japan\\
$^{2}$Advanced Meson Science Laboratory, RIKEN Nishina Center, Wako 351-0198, Japan\\
$^{3}$Department of Materials Science and Engineering,
Kyoto University, Kyoto 606-8501, Japan}



\date{\today}

\begin{abstract}
$^{121/123}$Sb nuclear quadrupole resonance and muon spin relaxation experiments of Mo$_3$Sb$_7$ revealed symmetry breakdown to a nonmagnetic state below the transition recently found at $T_{\rm S} \simeq 50$ K.
The transition is characterized by a distinct lattice dynamics suggested from narrowing of nuclear fields.
We point out that the Mo sublatice is a unique three-dimensional frustrated lattice where nearest-neighbor and next-nearest-neighbor antiferromagnetic interactions compete, and propose that tetragonal distortion to release the frustration stabilizes long-range order of spin-singlet dimers, i.e., valence bond crystal, which is thermally excited to the dynamic state with cubic symmetry. 

\end{abstract}

\maketitle

Exotic phenomena induced by the geometric frustration have attracted great interest, and trial to find novel phases in various types of frustrated lattices has been going on \cite{review}. 
In the case of three-dimensional (3D) pyrochlore lattice, the symmetry breakdown to spin-Peierls-like states or to the valence bond crystal with the quenched spin degrees of freedom \cite{yamashita, tchernyshyov} is one of the options to release the frustration; a typical example is found in the case of MgTi$_2$O$_4$ \cite{isobe}. 
On the other hand, in the field of correlated-electron superconductivity (SC), another phase transition prior to SC, such as the antiferromagnetic (AF) or hidden order \cite{settai_amitsuka}, attracts interest, because the particular electronic states may be a key to understand the unconventional SC.

\begin{figure}[tb]
\includegraphics[width = 6.0cm]{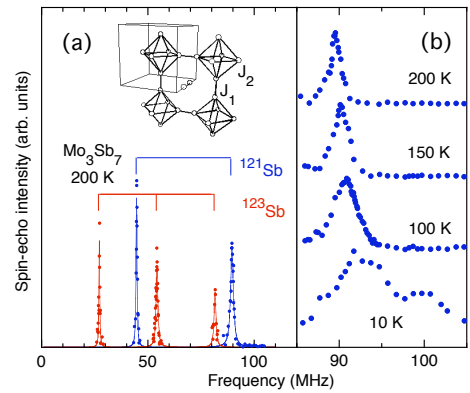}
\caption{\label{fig1}
(color online).
(a) $^{121/123}$Sb NQR spectrum of Mo$_3$Sb$_7$ at 200 K. 
Inset shows the Mo sublattice.
(b) Temperature variation of $^{121}$Sb-$2\nu_{\rm Q}$ line. 
}
\end{figure}

A metallic compound Mo$_3$Sb$_7$, with the cubic Ir$_3$Ge$_7$ type structure (space group $Im\bar{3}m$), is a superconductor with a critical temperature $T_{\rm c} \simeq 2.3$ K \cite{bukowski}. 
As seen in the inset of Fig.\ \ref{fig1}(a), the Mo sublattice ($12e$) is the 3D network of Mo-Mo dumbbells formed by nearest neighbors (NN).
Next-nearest-neighbor (NNN) bonds form an octahedral cage at the body-centered position.
The sufficiently short NN distance (3.0~\AA) comparing with the NNN distance (4.6~\AA) makes us anticipate easy dimerization between NN pairs.
Recently, Candolfi {\it et al.} \cite{candolfi} regarded Mo$_3$Sb$_7$ as an itinerant electron system and suggested coexistence of SC and spin fluctuations, which is a rare example in d electron metals. 
A broad maximum in the temperature ($T$) dependence of the susceptibility ($\chi$) with relatively large magnitude, a large electronic specific heat coefficient, a $T^2$ dependence of low-$T$ electrical resistivity, the Kadowaki-Wood relation between the specific heat and the resistivity, {\it etc.} point to the presence of strong electron correlation. 
Band structure calculations for cubic Mo$_3$Sb$_7$ \cite{dashjav, tran2}, showing appreciable contribution of Mo-4d orbitals to the density of states near the Fermi level, are in agreement with these observations.
Although no phase transition other than that at $T_{\rm c}$ was known in \cite{candolfi}, Tran {\it et al.} \cite{tran} reported recently the presence of another phase transition at $T^* \simeq 50$ K (in this article, referred to as $T_{\rm S}$), and proposed spin gap formation with the magnitude of 120 K associated with Mo-Mo dimerizaiton. 
Here the Mo$^{5+}$ state with spin $S = 1/2$ was presumed implicitly.
In fact at sufficiently high $T$, $\chi$ shows a $S = 1/2$ Curie-Weiss behavior with a large negative Weiss constant, suggesting strong AF interaction among active Mo spins.
Thus the Mo electronic state is of interest as the system with both itinerant and localized characters.
Furthermore Mo$_3$Sb$_7$ is now recognized as a new material with another phase transition prior to SC.
As for SC, the pairing mechanism is still controversial. 
Although most studies point to a BCS-type conventional SC \cite{bukowski, dmitriev1, candolfi2, tran2}, unconventional pairing has also been discussed  \cite{dmitriev2, candolfi2}. 
In \cite{tran}, it was argued that the Fermi surface nesting leads to SC. 

In this article, to manifest microscopic and dynamic aspects of the magnetism in Mo$_3$Sb$_7$, particularly to identify the phase between $T_{\rm c}$ and $T_{\rm S}$, we present $^{121/123}$Sb nuclear quadrupole resonance (NQR) and muon spin relaxation ($\mu$SR) results together with low-$T$ x-ray diffraction (XRD) data. 
We discuss possible frustration in the Mo sublattice by assuming AF NN and NNN interactions, and propose new exotic states realized due to the competition between NN and NNN interactions. 

Polycrystalline Mo$_3$Sb$_7$ specimen were prepared via reaction between liquid Sb and solid Mo. 
NQR spin-echo measurements were carried out using a conventional phase-coherent-type pulse spectrometer at 1.4--200 K.
Zero-field (ZF) $\mu$SR measurements were made at RIKEN-RAL Muon Facility at the Rutherford-Appleton Laboratory in the UK using a pulsed positive surface muon beam at 5--250 K. 
Low-$T$ powder XRD was measured using a conventional diffractmeter (MacScience MXP) with Cu $K\alpha$ radiation. 

Figure \ref{fig1}(a) shows the $^{121/123}$Sb NQR spectrum of Mo$_3$Sb$_7$ at 200 K. 
Antimony has two isotopes $^{121}$Sb (nuclear spin $I = 5/2$) and $^{123}$Sb ($I = 7/2$), which yield two and three quadrupole transition lines, respectively, when the electric field gradient (EFG) is axially symmetric. 
In cubic Mo$_3$Sb$_7$, there are two nonequivalent Sb positions with nonaxial and axial symmetries ($12d$ and $16f$, respectively). 
We have found only five lines at 200 K corresponding to either site. 
This is probably because the nuclear spin relaxation rate at the unobserved site is enhanced due to fast fluctuations of active Mo electron spins. 
The frequency intervals between neighboring lines of $^{121/123}$Sb are approximately equal, indicating nearly axial EFG at the observed site. 
$^{121/123}$Sb quadrupole frequencies are $^{121/123}\nu_{\rm Q} \simeq 47.7/27.2$ MHz at 200 K. 

\begin{figure}[tb]
\includegraphics[width = 6.5cm]{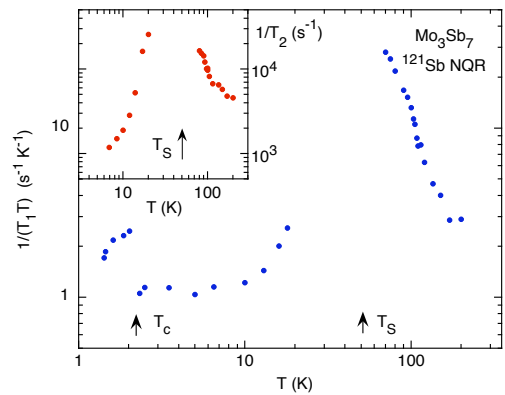}
\caption{\label{fig2}
(color online).
Temperature dependence of $^{121}$Sb-$1/(T_1T)$ for Mo$_3$Sb$_7$. Inset shows the temperature dependence of $1/T_2$.
}
\end{figure}

The $T$ dependence of the $^{121}$Sb-$2\nu_{\rm_Q}$ line (transitions of $| \pm 3/2 \rangle \leftrightarrow |\pm 5/2 \rangle$) is shown in Fig.\ \ref{fig1}(b).
The line is shifted to the high frequency side and gradually broadened with approaching $T_{\rm S}$. 
At $\sim T_{\rm S}$, the line loses intensity and disappears due to highly enhanced spin-echo decay rate $1/T_2$ (see the inset of Fig.\ \ref{fig2}). 
Below 20 K the signal reappears with complicated structures (see a part of the data at 10 K in Fig.\ \ref{fig1}(b)). 
Actually ZF nuclear resonance signal was continuously observed in almost the whole frequency range below 110 MHz with a number of peaks. 
This indicates clearly symmetry breakdown below $T_{\rm S}$. 
The origin is either appearance of the internal field (due to magnetic order), separation of crystallographic Sb sites (due to structural deformation) or both. 
Unfortunately, the spectrum at low $T$ is too complicated and unresolved to conclude this issue because of line split together with broadening, as well as the possible emergence of the signal from another Sb site. 
An early $^{121}$Sb M\"ossbauer study is not articulate either \cite{donaldson}. 
Hence, to determine the magnetic ground state, we employ another microscopic probe, $\mu$SR, which is sensitive to the appearance of the internal field; the result will be shown below.

The nuclear spin-lattice relaxation rate $1/T_1$ was measured for both $^{121}$Sb- and $^{123}$Sb-$2\nu{\rm_Q}$ transitions. 
Nuclear magnetization recovery was fitted well using the multi-exponential function with a single parameter expected for the case of axial EFG \cite{maclaughlin, note_T1}. 
The ratio of relaxation rates for the different isotopes was estimated to be $^{123}T_1/^{121}T_1 \simeq 3.5$ at $T > T_{\rm S}$, which is close to the squared gyromagnetic ratios $^{121}\gamma_n^2/^{123}\gamma_n^2 \simeq 3.4$. 
This confirms that the relaxation process is not by fluctuations of EFG but by those of the internal magnetic field at least above $T_{\rm S}$. 
$1/T_1$ was measured at the peak position of the $^{121}$Sb-$2\nu_{\rm_Q}$ line and plotted in the form of $1/(T_1T)$ vs.\ $T$ in Fig.\ \ref{fig2}. 
The divergent behavior was observed at $\sim T_{\rm S}$, although reliable values of $1/T_1$ could not be obtained near $T_{\rm S}$ due to poor signal-to-noise ratio \cite{note_T1T}. 
In fact, the transition at $T_{\rm S}$ is characterized by the stronger $T$ dependence of $1/T_2$ (the inset of Fig.\ \ref{fig2}); $1/T_2 \sim 10$--$10^3(1/T_1)$.
Below $T_{\rm S}$, $1/(T_1T)$ decreases rapidly with decreasing $T$, being consistent with the possible existence of spin gap \cite{tran}, but turns to show a $T_1T = constant$ Korringa behavior below $\sim 10$ K, indicating the presence of residual density of states at the Fermi level, which is related with the residual $\chi$ at low $T$. 
$1/T_1$ is enhanced just below $T_{\rm c}$. 
Such an enhancement is known as a coherence peak and usually observed for the conventional BCS superconductor with an isotropic gap \cite{hebel}.

\begin{figure}[tb]
\includegraphics[width = 7.0cm]{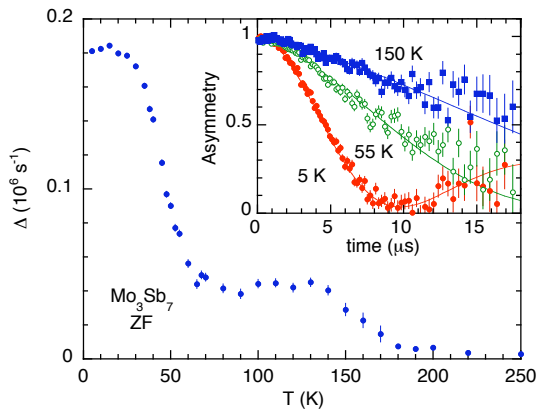}
\caption{\label{fig3}
(color online).
Temperature dependence of the ZF Kubo-Toyabe relaxation rate $\Delta$ in Mo$_3$Sb$_7$. 
Inset shows typical examples of ZF-$\mu$SR spectra (at 5, 55 and 150 K). Solid curves indicate the fit by the damped KT function.
}
\end{figure}

\begin{figure}[tb]
\includegraphics[width = 7.0cm]{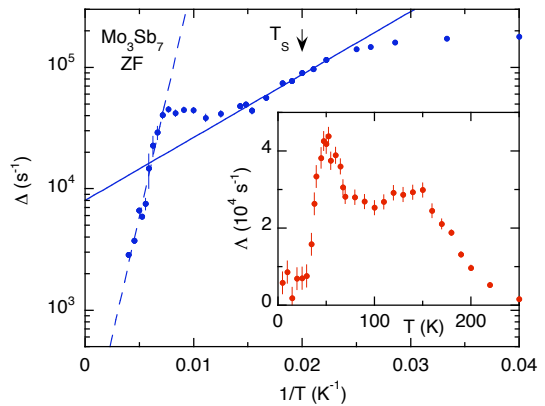}
\caption{\label{fig4}
(color online).
ZF Kubo-Toyabe relaxation rate $\Delta$ in Mo$_3$Sb$_7$ plotted against $1/T$. 
Activation energies for two different thermal excitation processes are estimated to be 120 and 910 K from the solid and broken lines, respectively. 
Inset shows the temperature dependence of the relaxation rate $\Lambda$.
}
\end{figure}

Now we shift to $\mu$SR results.
Typical examples of ZF-$\mu$SR spectra are shown in the inset of Fig.\ \ref{fig3}. 
Even at the lowest $T$ (5 K), no straightforward evidence of magnetic order such as muon spin precession was observed. 
At all $T$, the feature of relaxation is essentially the same and shows Gaussian-like depolarization.
The commonly used damped Kubo-Toyabe (KT) function
$P_\mu(t) = \exp(-\Lambda t) G_z^{\rm KT}(\Delta, t)$
with
$ G_z^{\rm KT} (\Delta, t) = \frac{1}{3} + \frac{2}{3} (1 - \Delta^2 t^2) \exp (-\frac{1}{2} \Delta^2 t^2) $ 
was fit to the data, where $\Delta/\gamma_{\mu}$ is the width of the static field distribution ($\gamma_{\mu}$ the muon gyromagnetic ratio), and $\Lambda$ is the damping rate associated with an additional relaxation process. 
As seen below, it is likely that the KT term is dominated by the nuclear dipolar field and the damping part reflects the Mo electronic system.
Figure \ref{fig3} shows the $T$ dependence of $\Delta$. First, let us see the $T$-independent part in the intermediate $T$ range of $\sim 70$--140 K with $\Delta/\gamma_{\mu} \simeq 0.25$ G. 
This tiny field is most probably due to dipolar fields coming from randomly oriented nuclear spins. 
Above $\sim 140$ K, $\Delta$ is reduced markedly. 
As shown in Fig.\ \ref{fig4}, $\ln \Delta$ is in proportion to $1/T$ in the $T$ range, suggesting a motional narrowing process caused by muon hopping among different sites, that is frequently observed at high $T$; the activation energy is $E_{\rm a}/k_{\rm B} \simeq 910$ K. 
With decreasing $T$ passing through $T_{\rm S}$, $\Delta$ is enhanced strongly as seen in Fig.\ \ref{fig3}. 
This behavior is clearly related with the transition at $T_{\rm S}$. 
The origin of the additional internal field below $T_{\rm S}$ is the most essential point in this experiment.
The Gaussian-type random distribution with small $\Delta/\gamma_{\mu} \simeq 2.1$ G even at the lowest $T$ is hardly attributed to the electron spin freezing but is reasonably explained as due to the change in nuclear dipolar field associated with the structural deformation mentioned below \cite{note_muSR}. 
The $T$ dependence of $\Delta$ at $\sim T_{\rm S}$ is again explained as a thermal excitation process with activation energy of $E_{\rm a}/k_{\rm B} \simeq 120$ K. 
This implies that the $T$ variation is dominated by a {\it dynamic} process, giving very fruitful information on the nature of the transition at $T_{\rm S}$.

The $T$ dependence of $\Lambda$ is shown in the inset of Fig.\ \ref{fig4}. 
The decrease above $\sim 150$ K would be analytical artifact due to the evlution of the relaxation process associated with the dynamic muon hopping.
$\Lambda$ has a sharp peak at $T_{\rm S}$ and decreases rapidly below $T_{\rm S}$. 
This behavior is qualitatively same as that of NQR-$1/T_1$, and reflects the phase transition in the Mo electronic system.
The quantitative difference in NQR-$1/T_1$ and $\mu$SR-$\Lambda$ is most probably due to the difference in the coupling mechanism; the interaction between Sb nuclei and Mo electrons is probably dominated by the direct hyperfine coupling, while fields from Mo electrons to the muon site is by the dipolar mechanism.
Different time windows of the measurements would also be appreciable.
Furthermore, the muon stopping site may be a high symmetric position, say the center of the Mo cage, where the fields from Mo electron spins would mostly be canceled.

\begin{figure}[tb]
\includegraphics[width = 7cm]{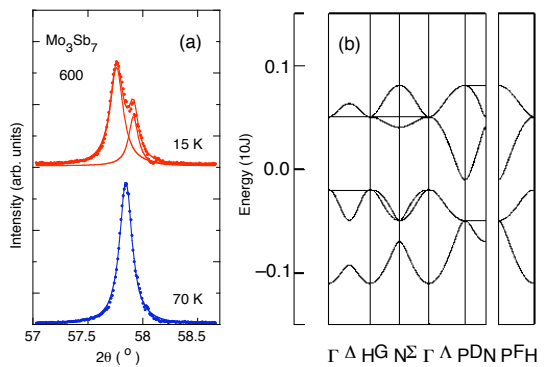}
\caption{\label{fig5}
(color online).
(a) X-ray 600 peak of Mo$_3$Sb$_7$ at 70 K and its split pattern at 15 K \cite{note_xray}. 
The contribution of $K\alpha_2$ radiation is subtracted by profile fitting. 
 (b) Dispersion curves of the exchange integral $J(q)$ of the Mo sublattice assuming negative $J_1$ and $J_2$. $J_1 = -0.05$ and $J_2 = 0.3J_1$ were used arbitrarily.
}
\end{figure}

The $\mu$SR result points to the nonmagnetic ground state. This is in accordance with the spin-singlet dimerization suggested in \cite{tran}. 
To get information on the lattice symmetry below $T_{\rm S}$, we made conventional XRD analyses at low $T$ and found appreciable difference in the patterns below and above $T_{\rm S}$. 
To see the change efficiently, we particularly show 600 diffraction peaks at 70 and 15 K in Fig.\ \ref{fig5}(a) \cite{note_xray}. 
As seen in the figure, a high-$T$ singlet splits into double peaks with the intensity ratio of $\sim 2 : 1$ by crossing $T_{\rm S}$, suggesting structural deformation from cubic to tetragonal (space group $I4/mmm$) as the simplest possibility; the estimated lattice parameter $a$ is larger than $c$ by $\sim 0.2$\% at 15 K. 

Although Mo$_3$Sb$_7$ is a metallic magnet, here we focus on the local nature of the Mo electron.
Taking a glance at the Mo sublattice, if the NN interaction $J_1$ is dominant, no substantial symmetry breakdown is necessary to form dimers but isotropic shrinkage is sufficient. 
Therefore in discussing the mechanism of the transition, the low-$T$ lattice symmetry is essential and NNN interaction $J_2$ should be taken into account. 
Note that positive (ferromagnetic) $J_2$ as suggested in \cite{tran} would not destabilize the cubic state. 
Here assuming negative $J_1$ and $J_2$, we calculated eigenvalues of their Fourier transform $J(q)$ for the cubic state in the scheme of the LCAO (linear combination of atomic orbitals) approximation; see Fig.\ \ref{fig5}(b). 
Further neighbor interactions are neglected for simplicity.
A flat dispersion of $J(q)$ found along the highest P-(D)-N branch indicates macroscopic degeneracies of AF ground states \cite{harris}; the Mo sublattice is not fully frustrated like pyrochlore but is apparently frustrated for all AF states with ${\bf q} = (\frac{1}{2} \frac{1}{2} \xi)$ with arbitrary $\xi$. 
We also found that the suggested tetragonal deformation lifts the degeneracies.
Thus we conclude that the cubic structure is unstable for negative $J_2$. 
The lattice deformation may be spin-frustration-driven symmetry breakdown like the spin Jahn-Teller (JT) effect \cite{yamashita}. 
In Mo$_3$Sb$_7$, the spin frustration triggers the tetragonal deformation to decouple the 3D isotropic AF interaction, and stabilizes spatial long-range order of spin-singlet dimers, i.e., the valence bond crystal.
As the reduced $\chi$ from higher $T$ ($\gg T_{\rm S}$) and the considerably broadened specific heat ($C$) anomaly were reported \cite{tran}, this state is realized instead of long-range AF order due to strong intra-dimer coupling already developed from sufficient high $T$. 

Taking account of the motional narrowing of nuclear dipolar fields observed by $\mu$SR at $\sim T_{\rm S}$, and structural characteristics of the low-$T$ phase, we propose a scenario for the transition.
First, note that the deformed states along [100], [010] and [001] directions are degenerate. 
Here we refer to the corresponding Mo electronic states as $\psi_x$, $\psi_y$ and $\psi_z$. 
We expect that the high-$T$ cubic state is the dynamic mixing of the three states, i.e., $(\psi_x + \psi_y + \psi_z)/3$, just like the vibronic dynamic JT state. 
In other words, harmonic oscillations of the valence bonds keeping strong intra-dimer interaction are thermally induced. 
The narrowing of $\mu$SR-$\Delta$, which reflects a relatively high-energy part of the dynamics, is explained naturally; at sufficiently low $T$, muons sense well defined nuclear dipolar fields from the deformed Mo sublattice, which are averaged by the dynamic atomic motion with increasing $T$. 
NQR-$1/T_2$ above $T_{\rm S}$, which nearly follows $\mu$SR-$\Delta$, is explained by the same mechanism, while the damping below $T_{\rm S}$ is ascribed to the decoupling of the nuclear spin-spin interaction, i.e., the reduction of Sb like-nuclei due to the separation of the Sb sites in the distorted state. 
It is of interest that magnetic entropy estimated from $C$ exceeds $R\ln 2 \simeq 0.69R$ corresponding to the $S = 1/2$ manifold and is asymptotic to a much higher value $\sim 1.0R$ \cite{tran} close to $R\ln 3 \simeq 1.1R$ expected for the triply degenerate state. 

In conclusion, NQR and $\mu$SR experiments revealed symmetry breakdown to a nonmagnetic state below $T_{\rm S} \simeq 50$ K, with most probably tetragonal lattice distortion.
Assuming negative $J_1$ and $J_2$ in the Mo sublattice, we pointed out possible frustration and proposed long-range order of spin-singlet dimers to the valence bond crystal, which transfers to the characteristic dynamic state with the cubic symmetry.
Precise structural determination and experiments to detect phonon anomalies would be helpful to discuss the mechanism of the transition. 
On the other hand, NQR-$1/T_1$ demonstrated the $s$-wave SC emerged from a metallic state where high-$T$ Mo spins are quenched.
The electron itinerancy and the present description focusing on the local nature should be reconciled in further studies, similarly to in other geometrically frustrated metals \cite{ballou}.

We acknowledge T. Mito and K. Ueda for helpful discussions, Y. Narumi and K. Kindo at ISSP, the University of Tokyo, and Y. Tanaka at RIKEN, SPring-8 for collaboration.
This work was supported by a grant from the University of Hyogo, and by Grant-in-Aid for Scientific Research on Priority Areas ``Novel States of Matter Induced by Frustration" (19052003).


\begin{thebibliography}{99}

\bibitem{review}
A. P. Ramirez, in {\it Handbook of Magnetic Materials}, edited by K. H. J. Buschow (North-Holland, Amsterdam, 2001).

\bibitem{yamashita} Y. Yamashita and K. Ueda, Phys. Rev. Lett. \textbf{85}, 4960 (2000).

\bibitem{tchernyshyov} 
O. Tchernyshyov {\it et al.}, Phys. Rev. B \textbf{66}, 064403 (2002). 

\bibitem{isobe} 
M. Isobe and Y. Ueda, J. Phys. Soc. Jpn. \textbf{71}, 1848 (2002), 
M. Schmidt {\it et al.}, Phys. Rev. Lett. \textbf{92}, 056402 (2004),
S. Di Matteo {\it et al.}, Phys. Rev. Lett. \textbf{93}, 077208 (2004).

\bibitem{settai_amitsuka} 
For example, 
R. Settai {\it et al.}, J. Phys. Soc. Jpn. \textbf{76}, 051003 (2007), 
H. Amitsuka {\it et al.}, Phys. Rev. Lett. \textbf{83}, 5114 (1999).

\bibitem{bukowski} 
Z. Bukowski {\it et al.}, Solid State Commun. \textbf{123}, 283 (2002).

\bibitem{candolfi} 
C. Candolfi {\it et al.}, Phys. Rev. Lett. \textbf{99}, 037006 (2007).

\bibitem{dashjav}
U. H\"aussermann {\it et al.}, Chem. Eur. J. \textbf{4}, 1007 (1998),
E. Dashjav {\it et al.}, J. Mater. Chem. \textbf{12}, 345 (2002)

\bibitem{tran2}
V. H. Tran {\it et al.}, arXiv: 0803.2948v1.

\bibitem{tran} 
V. H. Tran {\it et al.}, Phys. Rev. Lett. \textbf{100}, 137004 (2008).

\bibitem{dmitriev1}
V. M. Dmitriev {\it et al.}, Supercond. Sci. Technol. \textbf{19}, 573 (2006).

\bibitem{candolfi2} 
C. Candolfi {\it et al.}, Phys. Rev. B \textbf{77}, 092509 (2008).

\bibitem{dmitriev2} 
V. M. Dmitriev {\it et al.}, Low Temp. Phys. \textbf{33}, 295 (2007). 

\bibitem{donaldson}
J. D. Donaldson {\it et al.}, Acta Chem. Scand. A \textbf{28}, 866 (1974).

\bibitem{maclaughlin} 
D. E. MacLaughlin and J. D. Williamson, Phys. Rev. B \textbf{4}, 60 (1971).

\bibitem{note_T1}
Below $T_{\rm S}$, 1/$T_1$ was measured at the peak position near 90 MHz, and the  
assumption of axial EFG is only for simplicity.

\bibitem{note_T1T}
Above $T_{\rm S}$, $1/T_1$ varis roughly as $(T - T_{\rm S})^3$, suggesting unconventional electron-spin dynamics, presumably via electron-phonon coupling.

\bibitem{hebel} 
L. C. Hebel and C. P. Slichter, Phys. Rev. \textbf{113}, 1504 (1959).

\bibitem{note_muSR}
Strictly speaking, the possibility of magnetic order cannot be excluded, when the internal field from Mo electrons is canceled at the muon site.

\bibitem{note_xray}
In fact, the 600 reflection is superimposed with 442, of which intensity is smaller by one order. 

\bibitem{harris}
A. B. Harris {\it et al.}, Phys. Rev. B \textbf{45} (1992) 2899,
J. N. Reimers {\it et al.}, Phys. Rev. B \textbf{43} (1991) 865.

\bibitem{ballou}
R. Ballou {\it et al.}, Phys. Rev. Lett. \textbf{76}, 2125 (1996),
H. Nakamura  {\it et al.}, J. Phys.: Condens. Matter \textbf{9}, 4701 (1997),
S. Kondo {\it et al.}, Phys. Rev. Lett. \textbf{78}, 3729 (1997)

\end{thebibliography}

\end{document}